\documentclass[times, 10pt,twocolumn]{article} 
\usepackage{latex8}
\usepackage{times}

\usepackage{amssymb}
\usepackage{graphicx}
\usepackage{url}
\usepackage[vlined,ruled]{algorithm2e} 

\newcommand{\myscale}{0.48}

\newcommand{\noteperso}[1]{\begin{center}
 \fbox{\begin{minipage}{6cm}#1\end{minipage}}\end{center}}
\renewcommand{\noteperso}[1]{}

\newcommand{\axes}[3]{
\mbox{ }\hfill \scalebox{0.7}{\em #1} 
  \hfill \mbox{ } \hfill \scalebox{0.7}{\em #2} 
  \hfill \mbox{ } \hfill \scalebox{0.7}{\em #3} \hfill \mbox{ } \vskip -0.2cm}

\newcommand{\ie}{{\em i.e.}}
\newcommand{\ping}{{\tt ping}}
\newcommand{\traceroute}{{\tt trace\-route}}
\newcommand{\tracetree}{{\tt trace\-tree}}

\newcommand{\ip}{\mbox{\em \sc ip}}
\newcommand{\icmp}{{\em \sc icmp}}
\newcommand{\udp}{{\em \sc udp}}

\newcommand{\bfs}{{\em \sc bfs}}
\newcommand{\ttl}{\mbox{\sc ttl}}

\newcommand{\dsl}{\mbox{\sc dsl}}

\graphicspath{{Figures/}}

\begin{document}

\title{A Radar for the Internet}
\author{
Matthieu Latapy$\mbox{}^{1,2}$, Cl\'emence Magnien$\mbox{}^{1,2}$, Fr\'ed\'eric Ou\'edraogo$\mbox{}^{1,2,3}$\\
1: UPMC Univ Paris 06, UMR 7606, LIP6, F-75016, Paris, France\\
2: CNRS, UMR 7606, LIP6, F-75016, Paris, France\\
3: University of Ouagadougou, LTIC, Ouagadougou, Burkina Faso\\
\url{First-name.Lastname@lip6.fr}}

\maketitle
\thispagestyle{empty}

\begin{abstract}

  In contrast with most internet topology measurement res\-earch, our
  concern here is not to obtain a map as complete and precise as
  possible of the whole internet. Instead, we claim that each
  machine's view of this topology, which we call ego-centered view, is
  an object worth of study in itself. We design and implement an
  ego-centered measurement tool, and perform radar-like measurements
  consisting of repeated measurements of such views of the internet topology. We
  conduct long-term (several weeks) and high-speed (one round every
  few minutes) measurements of this kind from more than one hundred
  monitors, and we provide the obtained data. We also show that these
  data may be used to detect events in the dynamics of internet
  topology.

\end{abstract}

\section{Introduction.}
\label{sec_introduction}

Since the end of the nineties,
constructing maps of the internet using \traceroute -like measurements
received much attention, see for instance
\cite{faloutsos99sigcomm, dimes,skitter,trahome,scamper,ripeNccTtm,nlanrAmp,whatToDo,atlas,heuristics,scriptroute}.
Such measurements are however partial and they may contain significant
bias
\cite{sampling,marginal,plrevisited,dallAsta,relevance,achlioptas05bias}.
As a consequence, much effort is nowadays devoted to the collection of
more accurate data \cite{dimes,trahome,e2emon2007,parisTraceroute}, but
this task is challenging.

In order to avoid these issues and obtain some insight on internet topology {\em dynamics}, 
we use here a radically different approach: we focus on what a given machine sees of the topology around itself, which we call an {\em ego-centered view} (it basically is a routing tree measured in a \traceroute-like manner). 
These ego-centered measurements may be performed very efficiently (typically in minutes, and inducing low network load); it is therefore possible to repeat them in periodic rounds, and obtain in this way information on the {\em dynamics} of the topology, at a time-scale significantly higher than previous approaches (see for instance \cite{oliveira2007evol,plrevisited}).

Taking advantage of these strengths, we conduct massive radar-like measurements of the internet. We provide both the measurement tool and the collected data, and show that they reveal interesting features of the observed topology.

\section{Measurement framework.}
\label{sec_tools}

One may use \traceroute\ directly to collect ego-cent\-ered views by probing a set of destinations.
This approach however has serious drawbacks. First, as detailed in
\cite{DTSigmetrics} and illustrated in Figure~\ref{fig_ex}, the
measurement load is highly unbalanced between nodes and there is much
redundancy in the obtained data (intuitively, one probes links close
to the monitor much more than others). Even worse, this implies that
the obtained information is not homogeneous, and thus much more
difficult to analyse rigorously (for instance, the dynamics may seem
higher close to the monitor). Finally, though the measurement would
intuitively produce a routing tree, the obtained view actually differs
significantly from a tree (see for instance \cite{parisTraceroute}). Again,
this makes the analysis (visualisation of the data, for instance) more
intricate.

Finally, the direct \traceroute\ approach has multiple severe
drawbacks. In this section we first design an ego-centered measurement
tool remedying to this. We then include it in a radar measurement
scheme.

\subsection{Ego-centered measurements.}
\label{sec_tracetree}

\begin{figure}[!h]
\centering
\includegraphics[scale=0.27]{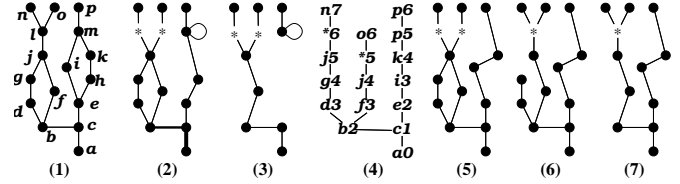}
\caption{Typical outputs of various measurements schemes.
(1) -- Real topology. 
$a$ is the monitor, $n$, $o$, and $p$ are the destinations.
We suppose that $l$ does not answer to probes, that $b$ is a per-destination load balancer,
forwarding traffic for $n$ to $d$,
and traffic for $o$ to $f$, and that $e$ is a per-packet load balancer forwarding packets alternately to $i$ and $h$. Such situations are frequent in practice.
(2) -- Measurement with \traceroute. Three routes are collected, leading to a higher load on links close to the monitor (represented by thicker lines here).
(3) -- Naive tree measurement. Because of a route change due to per-packet load balancer $e$, one obtains a disconnected part.
(4) -- Measurement with \tracetree. Nodes are pairs of \ip\ addresses and \ttl, with redundancy in the addresses;
one necessarily obtains a tree.
(5--7) -- Main steps of the filtering process.
(5) -- Pairs with same \ip\ address are merged and loops are removed;
(6) -- Appropriate stars are merged and a \bfs\ tree is computed;
(7) -- Leaves which are not the last node on a path towards a destination are iteratively removed.
This is the final output of the filter.
}
\label{fig_ex}
\end{figure}

As already discussed in various contexts
\cite{DTJSAC,DTSigmetrics,probingScheme,pansiot2007multicast,streamlining,scriptroute},
one may avoid the issues described
above by performing tree-like measurements in a backward way: given a
set of destinations to probe, one first discovers the last link on the
path to each of them, then the previous link on each of these paths,
and so on; when two (or more) paths reach the same node then the
probing towards all corresponding destinations, except one,
stops\,\footnote{Such measurements require the distance towards each
  destination, which is not trivial \cite{streamlining}; we discuss
  this in Section~\ref{sec_radar}.}.  However, as illustrated in
Figure~\ref{fig_ex}, naive such measurements encounter serious
problems because of routing changes and other events.
We provide a solution in
the \tracetree\ algorithm below: the tree nodes are not \ip\ addresses
anymore, but pairs composed of an \ip\ address (or a
star if a timeout occurred) and the \ttl\ at which it was observed
(see Figure~\ref{fig_ex} for an illustration). This is sufficient to
ensure that the obtained view is a tree, while keeping the algorithm
very simple. It sends only one packet for each link, and
thus is optimal.
Moreover, each link is discovered exactly
once, which gives an homogeneous view of the topology and balances the
measurement load.

\begin{algorithm}[H]
\caption{\tracetree\ algorithm.}
\SetCommentSty{small}
\dontprintsemicolon
\SetKw{Or}{or}
\SetKw{KwIf}{if}
\SetKw{Output}{print}
\SetKw{Gets}{$\leftarrow$\ }
\KwIn{set $D$ of destinations, with $d\in D$ at distance $ttl_d$.}
$to\_probe$ \Gets empty queue,
$to\_receive$ \Gets $\emptyset$,
$seen$ \Gets $\emptyset$\;
\lForEach{$d \in D$}{add $(d,ttl_d)$ to $to\_probe$}\;
\While{$to\_probe$ not empty \Or $to\_receive \not= \emptyset$}{
 \lnlset{send_strategy}{$\alpha$}%
 \If{$to\_probe$ not empty}{
  pop $(d,ttl)$ from $to\_probe$ and send a probe to it\;
  add $(d,ttl,current\_time())$ to $to\_receive$\;
  }
 \tcp{here necessarily $to\_receive \not= \emptyset$}
 \lnlset{check_packet}{$\beta$}%
 \If{answer $p$ to a probe to $(d,ttl)$ received}{
  \tcp{$p$ sent by $p.source$, reply to a probe to $(d,ttl)$}
  \If{$(d,ttl,\_) \in to\_receive$}{
    \tcp{else timeout}
    remove $(d,ttl,\_)$ from $to\_receive$;\\
   \Output $p.source$ $ttl$ $d$\;
   \If{$(p.source,ttl) \not\in seen$}{
    add $(p.source,ttl)$ to $seen$\;
    push $(d,ttl-1)$ in $to\_probe$ \KwIf $ttl>1$\;
    }
   }
  }
 \For{$(d,ttl,t) \in to\_receive$ \KwIf timeout exceeded}{
   remove $(d,ttl,t)$ from $to\_receive$\;
   \Output * $ttl$ $d$\;
   push $(d,ttl-1)$ in $to\_probe$ \KwIf $ttl>1$\;
  }
 }
\label{algo_tracetree}
\end{algorithm}

From such trees with (\ip,\ttl) nodes, one obtains a tree on \ip\ addresses by applying the following filter (illustrated in Figure~\ref{fig_ex})\,\footnote{The measurement would be slightly more efficient if the filter was included directly in \tracetree; however, to keep things simple and modular, we preferred to separate the two.}:
first merge all nodes of the tree which correspond to a same \ip;
remove loops (links from an \ip\ to itself); iteratively remove the stars with no successor;
merge all the stars which are successor of a same node into a unique star;
construct a \bfs\ tree of the obtained graph which leads to a tree on \ip\ addresses\,\footnote{During the construction
of the \bfs{} tree, neighbours of a node are visited in lexicographic order, and stars 
are visited after \ip{}s.};
iteratively remove the leaves which are not the last nodes encountered when probing any destination.

The key point is that the obtained tree is a possible \ip\ routing
tree from the monitor to the destinations (similar to a broadcast
tree).
The obtained tree contains almost as much information as the original
\tracetree\ output and has the advantage of being much more simple to
analyse.
We evaluated the impact of this filtering on our observations,
and found that it was negligible:
Detailing this is out of the scope of this paper.



Many non-trivial points would deserve more discussion. For instance,
one may apply a greedy sending or receiving strategy
(by replacing line $\alpha$ or $\beta$ in
Algorithm~\ref{algo_tracetree} by a {\tt while}, respectively);
identifying reply packets is
non-trivial, as well as extracting the relevant information from the
read packets; introducing a delay may be necessary to stay below the maximal
\icmp\ sending rate of the monitor; one may consider answers received
after the timeout but before the end of the measurement (whereas we
ignore them); one may use other protocols than \icmp\ (the classical
\traceroute\ uses \udp\ or \icmp\ packets); the initial order of the
destinations may have an impact on the measurement; there may be many
choices for the \bfs\ tree in the filter; etc.  However, entering in
such details is far beyond the scope of this paper, and we refer to
the code and its documentation \cite{radarurl} for full details.

\subsection{Radar.}
\label{sec_radar}

With the \tracetree\ tool and its filtered version, we have the ground
material to conduct radar measurements: given a monitor and a set of
destinations, it suffices to run periodic ego-centered measurements,
which we call measurement {\em rounds}.
The measurement frequency must be high enough to capture interesting
dynamics, but low enough to keep the network load reasonable. We will
discuss this 
in the next section.

The only remaining issue is the estimation of distances towards destinations, which is a non-trivial task in general \cite{streamlining}. This plays a key role here, since over-estimated distances lead to several packets hitting destinations. Under-estimated distances, instead, miss the last links towards the destinations.

One may however suppose that the distance between the monitor and any
destination generally is stable between consecutive rounds of radar
measurement.
Then,
the distances at a given round are the ones observed during the
previous round. If the distance happens to be under-estimated (we do
not see the destination at this distance), then we set it to a default
maximal value (generally equal to $30$) and start the measurement from
there (and we update the corresponding distance for the next round).

\begin{figure*}[!ht]
\centering
\includegraphics[scale=\myscale]{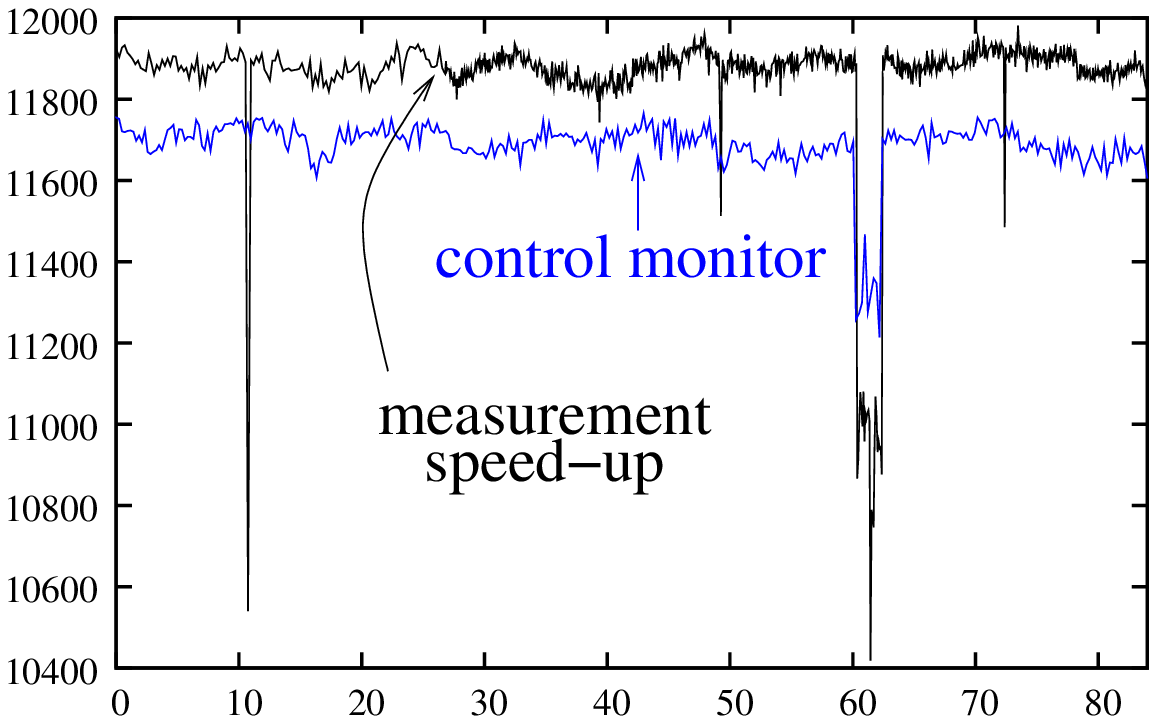}
\includegraphics[scale=\myscale]{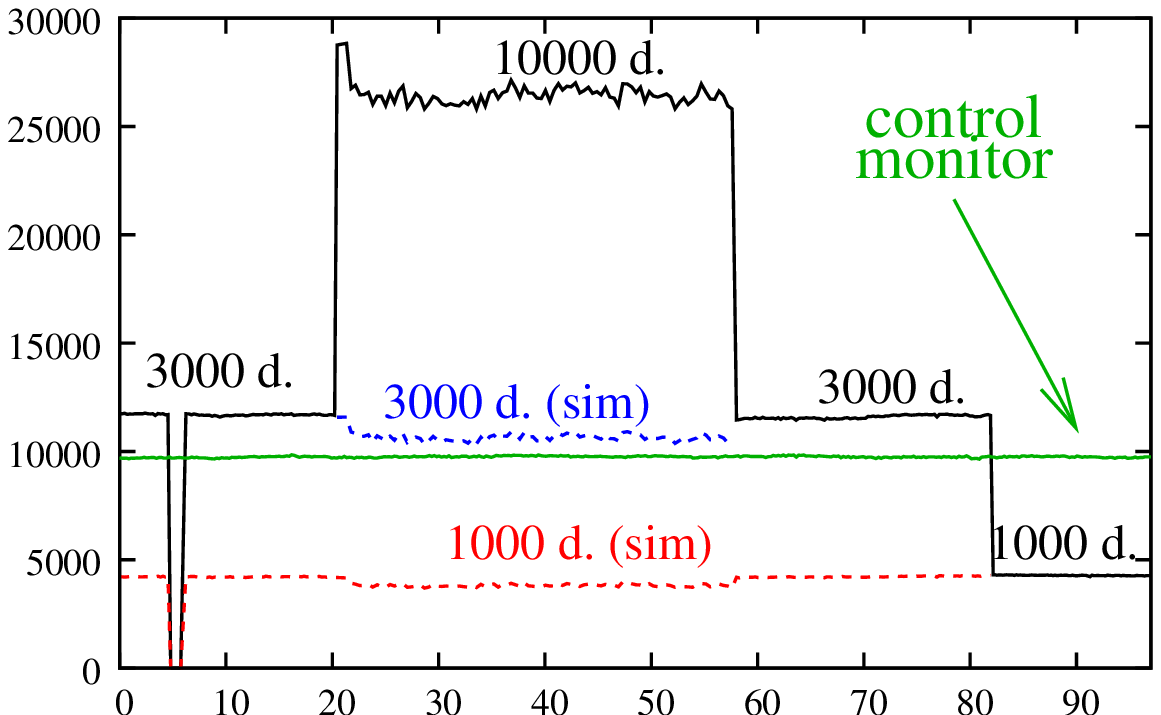}
\includegraphics[scale=\myscale]{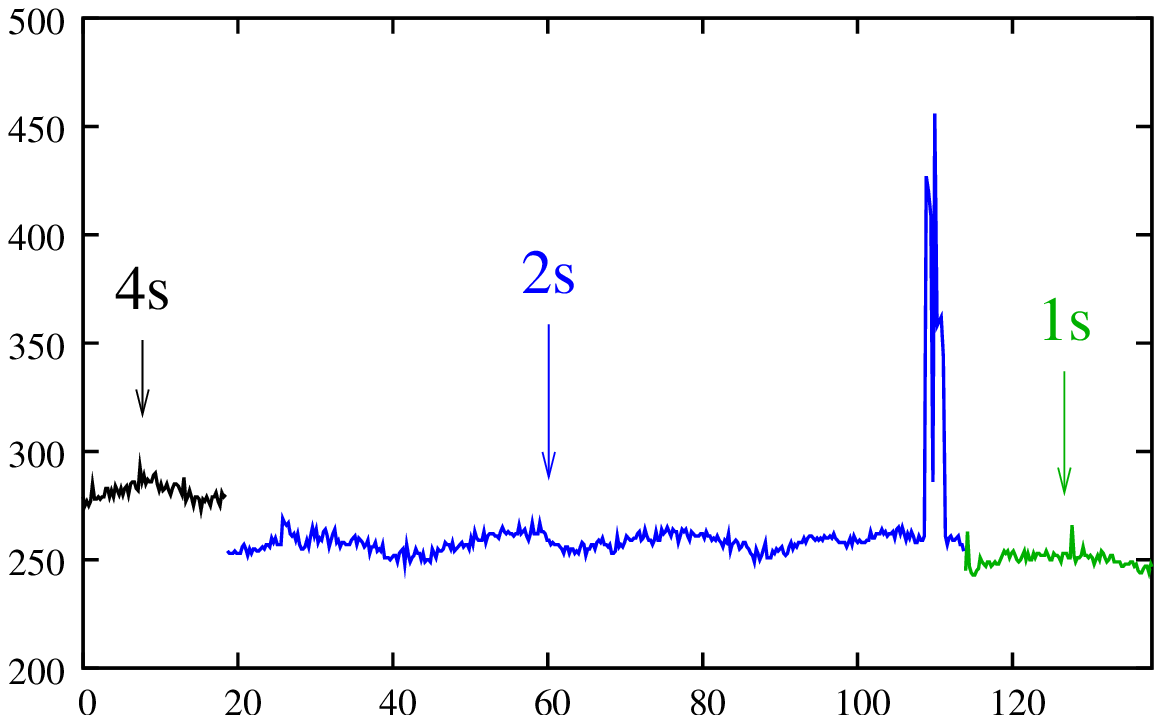}\\
\axes{\ \ \ \ \ \ \ \ \ \ \  \hspace{1cm}$x$ = hours; $y$ = \# ip}{\ \ \ \ \ \ \ \ \hspace{2.5cm} \em $x$ = hours; $y$ = \# ip}{\ \ \ \hspace{2cm} $x$ = hours; $y$ = round duration (s)}
\caption{
{\bf Impact of measurement parameters.} 
The $x$ axis of all plots represents the time (in hours) since the beginning of the measurement.
{\bf Left: impact of inter-round delay.} Number of distinct \ip{} addresses viewed at each round. The bottom plot corresponds to a control monitor with the base parameters; the other monitor starts with the base parameters, and about $27$ hours later we reduce the inter-round delay from $10$ minutes to $1$ (each ego-centered measurement takes around $4$ minutes).
{\bf Center: impact of the number of destinations.} Number of distinct \ip{} addresses viewed at each round.
The plot close to $y=10\,000$ corresponds to a control monitor with the base parameters. The other plain-line plot is produced by a monitor which starts with the base parameters, thus with a destination set $D$ of size $3\,000$, changes to a set $D'$ of $10\,000$ destinations containing $D$, 
goes back to $D$,
and finally turns to a subset $D''$ of size $1\,000$ of $D$.
In addition, the dotted plots are simulations of what we would have seen from this monitor with $D$ during the measurement using $D'$ (obtained by dropping all nodes and links which are on paths towards destinations that are not in $D$), and what we would have seen with $D''$ during the measurements using $D$ or $D'$ (obtained similarly).
{\bf Right: impact of timeout value.} Round duration (in seconds). 
The monitor starts with a timeout value of $4\,s$, then we change it to $2\,s$, and finally to $1\,s$.
}
\label{fig_speed_up}
\label{fig_nb_dest}
\end{figure*}

\section{Measurement and data.}
\label{sec_data}


First notice that many parameters (including the monitor and destination set) may have a deep impact on the obtained data. Estimating this impact is a challenging task since testing all combinations of parameters is totally out of reach. In addition, the continuous evolution of the measured object makes it difficult to compare several measurements: the observed changes may be due to parameter modifications or to actual changes in the topology.

To bypass these issues while keeping the study rigorous, we propose
the following approach. We first choose a set of seemingly reasonable
parameters, which we call {\em base parameters} (see
Section~\ref{sec_base}). Then we conduct measurements with these
parameters from several monitors in parallel.
On some monitors,
called {\em control monitors}, we keep these parameters constant;
on others, called {\em test monitors}, we alternate periods with base
parameters and periods where we change (generally one of) these
parameters.
Control monitors make it possible to check
that the changes observed from test monitors are due to changes of
parameters, not to events on the network. The alternation of periods
with base parameters and modified ones also makes it possible to
confirm this, and to observe the induced changes in the observations.
In many cases, it is also possible to simulate what one would have
seen in principle if the parameters had stayed unchanged, which gives
further insight (we will illustrate this below).

We use a wide set of more than one hundred monitors scattered around
the world, provided by PlanetLab \cite{planetLab} and other structures
(small companies and individual \dsl\ links) \cite{radarurl}.  In
order to be as general as possible, and to simplify the destination
setup, we use destinations chosen by sampling random valid \ip\
addresses and keeping those answering to \ping\ {\em at the time of
  the list construction}. Other selection procedures would of course
make sense (this raises interesting perspectives).

\subsection{Our base parameters and data set.}
\label{sec_base}

In all the paper, the base parameters consist of a set of $3\,000$ destinations for each monitor, a maximal \ttl\ of $30$, a $2$ seconds timeout and a $10$ minutes delay between rounds. All our measurements were conducted with variations of these parameters; wherever it is not explicitly specified, the parameters were the base ones. We ran measurements continuously during several weeks, with some interruptions due to monitors and/or local network shutdowns.
The obtained data is available at \cite{radarurl}.

\subsection{Influence of parameters.}
\label{sec_influence}

Using the methodology sketched above, we show here how to rigorously evaluate the influence of various parameters. We focus on a few representative ones only, the key conclusion being that the base parameters described above fit our needs very well.

Figure~\ref{fig_speed_up} (left) shows the impact of the inter-round delay: on the rightmost part the delay was significantly reduced, leading to an increase in the observation's time resolution (\ie\ more points per unit of time). It is clear from the figure that this has no significant impact on the observed behavior. In particular, the variations in the number of \ip\ addresses seen, though they have a higher resolution after the speed-up, are very similar before and after it. Moreover, the control monitor shows that the base time scale is relevant, since improving it does not reveal significantly higher dynamics.

Figure~\ref{fig_nb_dest} (middle) shows the impact of the number of
destinations. As expected, increasing this number leads to an increase
in the number of observed \ip\ addresses.
The key point however is that increasing
the number of destinations may lead to a relative loss of efficiency:
simulations of what we would have seen with $3\,000$ or $1\,000$
destinations display a smaller number of \ip\ addresses than direct
measurements with these numbers of destinations (the control monitor
proves that this is not due to a simultaneous topology change). This
is due to the fact that probing towards $10\,000$ destinations induces
too high a network load: since some routers answer to \icmp\ packets
with a limited rate only \cite{govindan2002estimating}, overloading
them makes them invisible to our measurements. Importantly, this does
not occur in simulations of $1\,000$ destination measurements from
ones with $3\,000$, thus showing that the load induced with $3\,000$
destinations is reasonable, to this regard.

Figure~\ref{fig_nb_dest} (right) shows the impact of the timeout value.
As expected, decreasing the timeout leads to a decrease in the round
duration.
However, it also causes more replies to probe packets to be ignored because we receive them after the timeout.
A good value for the timeout is a compromise
between the two. We observe that the round duration is only slightly
larger with a timeout of $2s$ than with a timeout of $1s$ (contrary to
the change between a timeout of $4$ and $2s$).  The base value of the
timeout (2s) seems therefore appropriate, because it is rather large
and does not lead to a long round duration.

\begin{figure*}[!ht]
\centering
\hfill
\includegraphics[scale=\myscale]{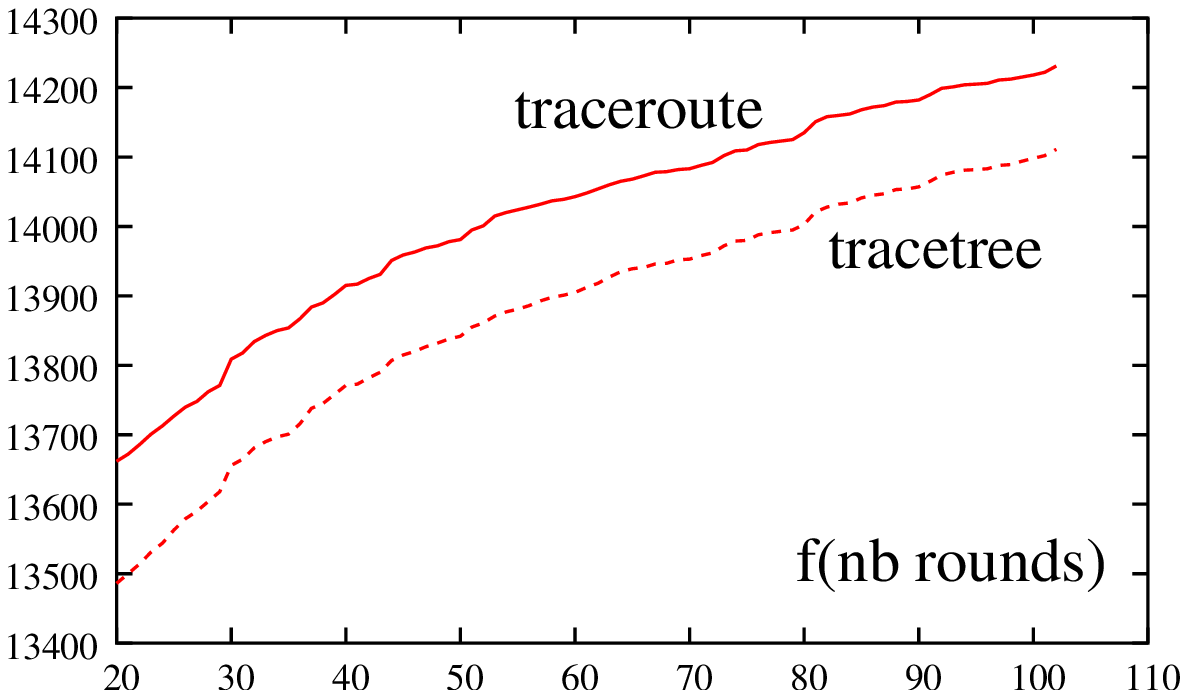}
\hfill
\includegraphics[scale=\myscale]{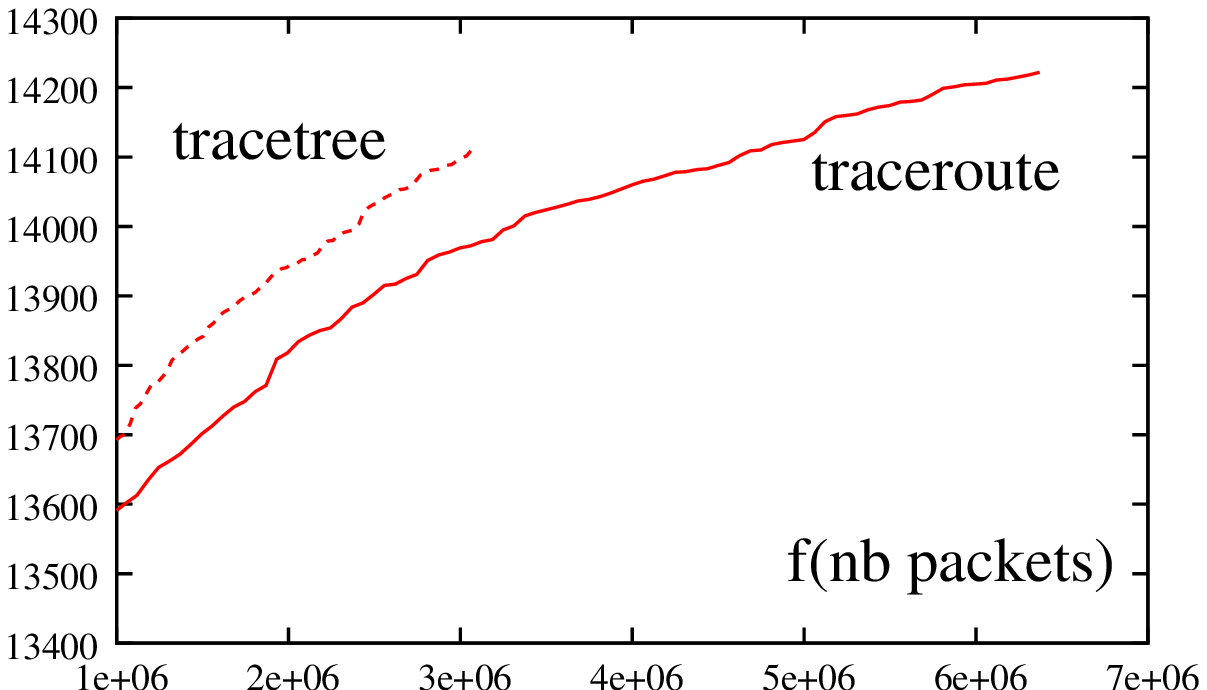}
\hfill
\includegraphics[scale=0.46]{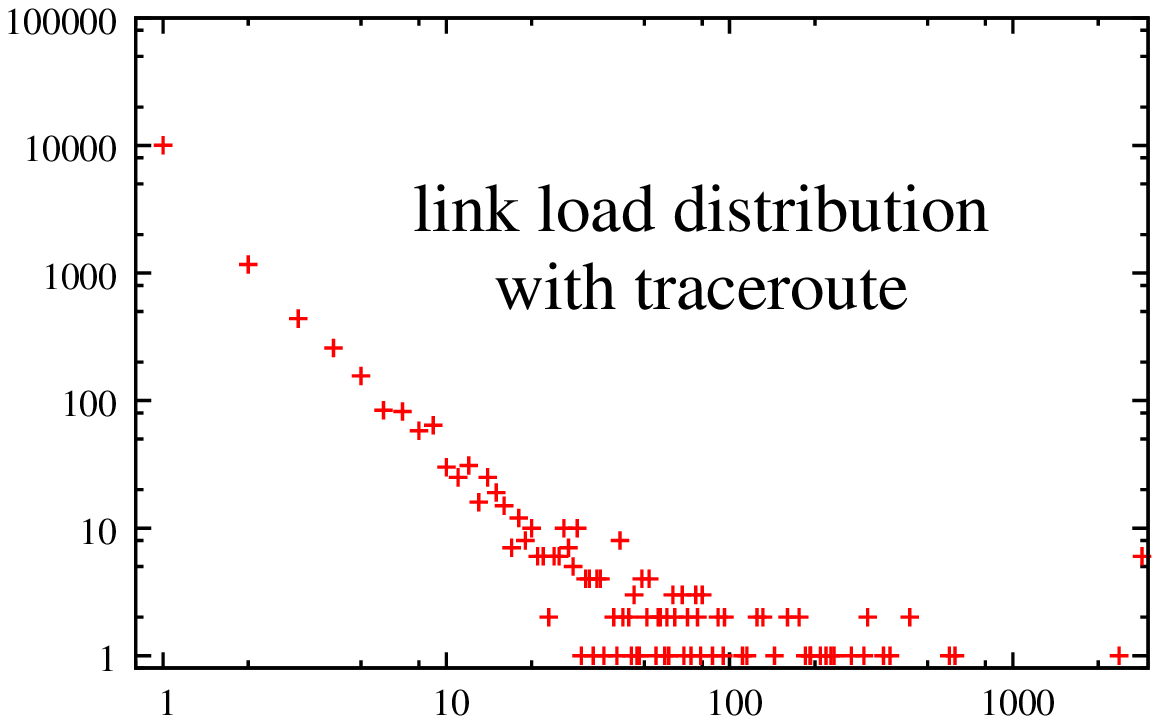}
\hfill
\\
\axes{\hspace{1cm}$x$ = \# rounds; $y$ = \# ip}{\hspace{0cm}$x$ = \# packets; $y$ = \# ip}{$x$ = \# times probed; $y$ = \# links\hspace{-1cm}}
\caption{
{\bf Comparison between \traceroute\ and \tracetree.}
{\bf Left and center:} number of distinct \ip\ addresses viewed since the beginning with a \traceroute\ measurement (plain lines) and a \tracetree\ measurement simulated from it (dotted lines);
left: as a function of the number of rounds; 
center: as a function of the number of packets sent.
To improve readability, we cut the part of the plots corresponding to the $20$ first rounds and to the $10^6$ first packets, respectively.
{\bf Right:} typical link load distribution with a \traceroute\ ego-centered measurement. For each value $x$ on the horizontal axis, we give the number of links which are discovered $x$ times during a \traceroute\ ego-centered measurement with 3\,000 destinations (base value).
}
\label{fig_traceroute}
\end{figure*}

We also considered other observables (like the number of stars seen at
each round, and the number of packets received after the timeout),
for measurements
obtained from various monitors and towards various destinations; in
all cases, the conclusion was the same: the base parameters proposed
above meet our requirements.

\begin{figure*}[!ht]
\centering
\includegraphics[scale=0.45]{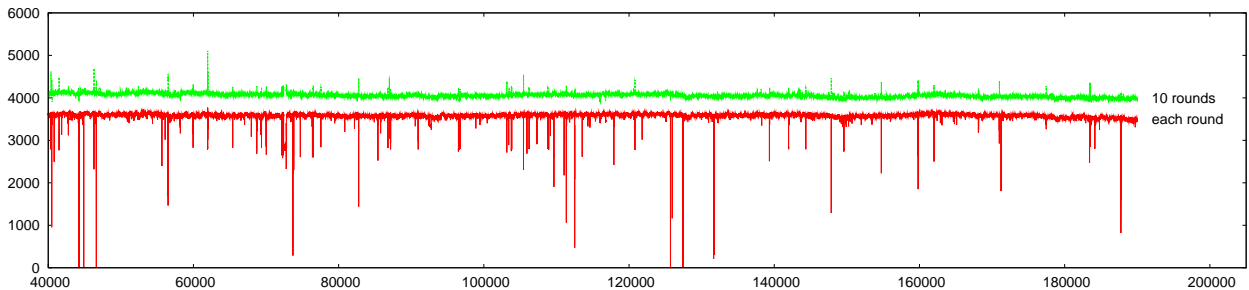}
\caption{Bottom: number of distinct \ip\ addresses observed during each round of measurement. 
Top: number of distinct \ip\ addresses observed during series of ten consecutive rounds.}
\label{fig_events}
\end{figure*}

\begin{figure*}[!ht]
\begin{minipage}[c]{0.35\linewidth}
\includegraphics[scale=0.48]{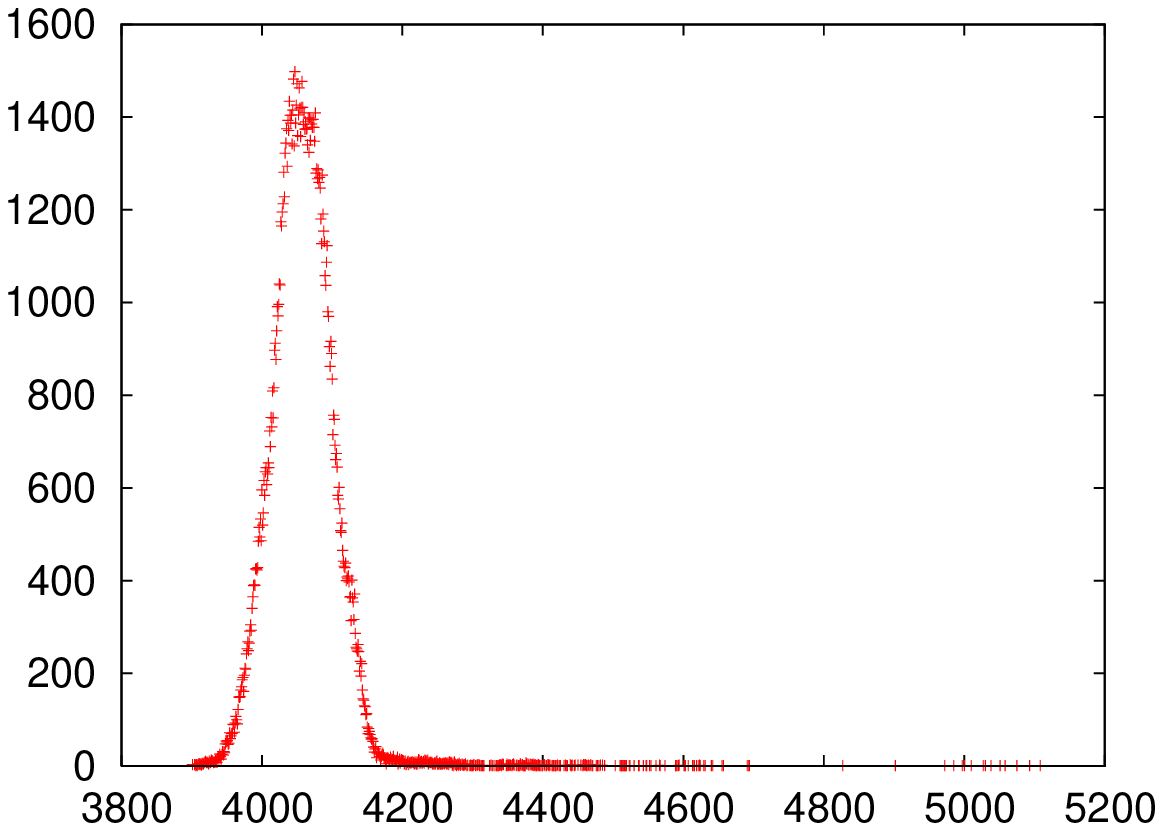}
\end{minipage} 
\begin{minipage}[c]{0.25\linewidth}
\includegraphics[scale=.28]{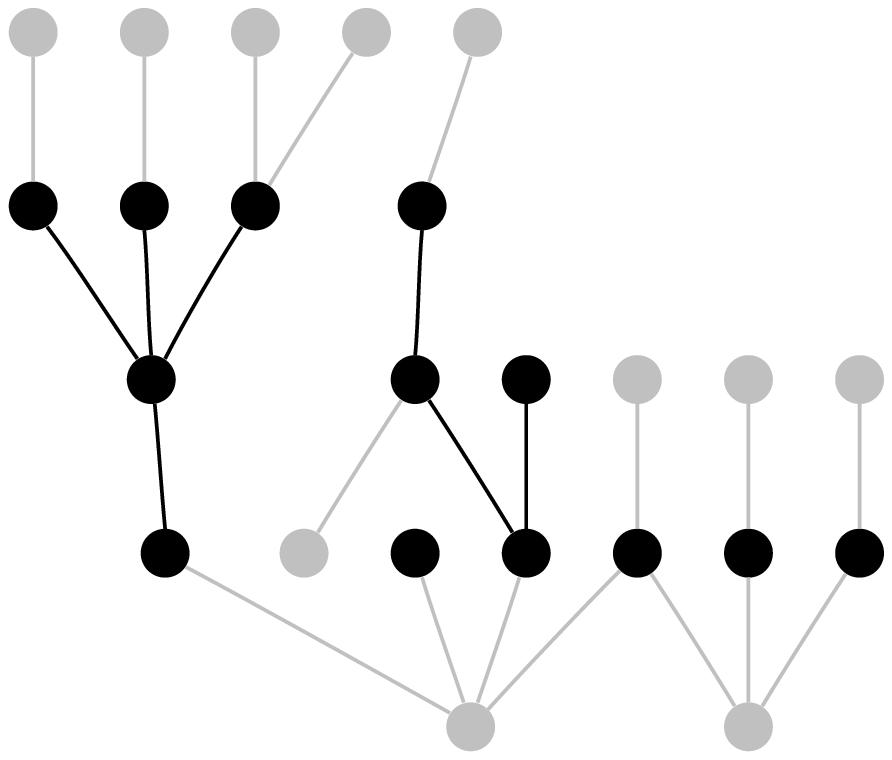}~
\includegraphics[scale=.28]{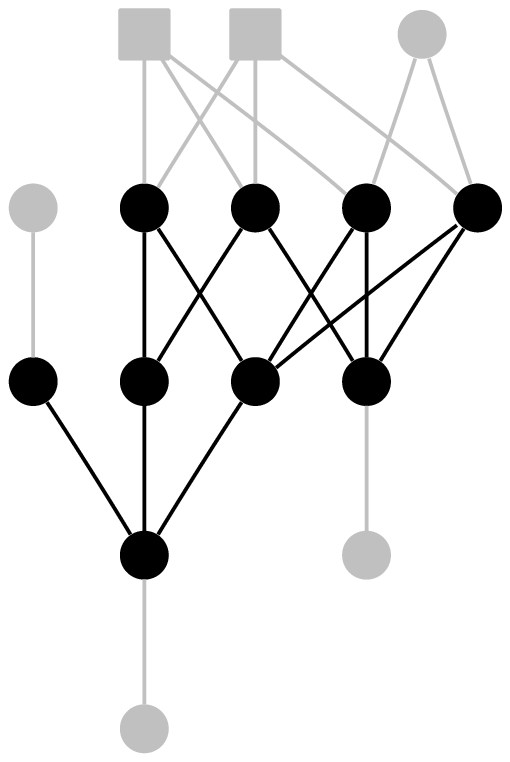}
\end{minipage} 
\begin{minipage}[c]{0.30\linewidth}
\includegraphics[scale=\myscale]{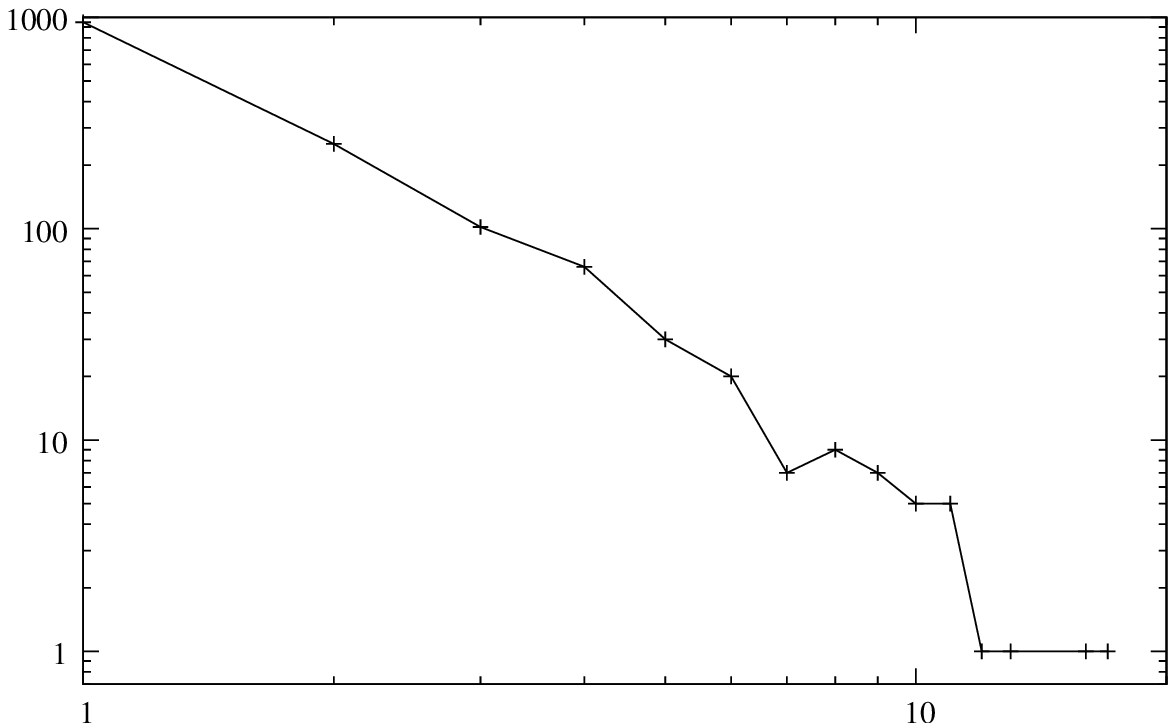}
\end{minipage}

\axes{\hspace{1cm}$x$ = \# ip ; $y$ = \# series}{\hspace{0cm}}{$x$ = components size; $y$ = \# components \hspace{-1cm}}
\caption{
{\bf Left: Distribution of the values of the upper plot in Figure~\ref{fig_events}}.
{\bf Center: typical {\em islands} of appearing nodes.} Each node is an \ip\ address; the black ones are the ones observed during the second half of the measurement only, the others being already present in the first half.
The square nodes were present in {\em all} the ($2\,200$) rounds of measurement. Links are directed from bottom to top, \ie\ from the monitor to destinations. The number of rounds necessary to discover all $13$ new nodes in the left drawing was $669$ rounds ($1\,306$ to $1\,974$), but only $2$ rounds ($2\,021$ and $2\,022$) were sufficient for the $9$ right ones. Notice that $7$ connected components of new nodes are displayed: $4$ of size $1$, $1$ of size $4$, $1$ of size $5$, and $1$ of size $9$.
{\bf Right: distribution of new node component sizes.} For each possible size $x$ (horizontal axis), the number of connected components of new nodes of size $x$ is given.
}
\label{fig_cc}
\end{figure*}

\subsection{Comparison with \traceroute.}
\label{sec_traceroute}

As explained in Section~\ref{sec_tracetree}, a key goal of our
\tracetree\ measurement tool is to perform significantly better than
direct use of \traceroute\ in our context.
To evaluate this, we compare the difference in the obtained information
with \traceroute{} and \tracetree{}, 
as well as the load they induce on the network (Figure~\ref{fig_traceroute}, left and center).
First notice that the plot as a function of the number of rounds with \traceroute\ is higher than the one with \tracetree, as expected: any \traceroute\ round gathers slightly more data than the corresponding \tracetree\ round (below $1\,\%$, here).
It is however much more interesting to compare them in terms of the number of packets sent (reflecting the load induced on the network and our ability to increase the measurement frequency). The plots show that, to this regard, \tracetree\ is much more efficient than direct \traceroute\ measurements: here, \tracetree\ reaches $14\,100$ distinct \ip\ addresses with around $3$ millions packets, while \traceroute\ needs around $4.5$ millions packets.

Recall moreover that the load induced by \tracetree\ is balanced among links,
which is not the case for \traceroute{}, see Figure~\ref{fig_traceroute} (right).
We can see that some links are probed a very high number of times {\em at each round}
(typically up to $3\,000$ times if we use $3\,000$ destinations).
See \cite{DTJSAC,DTSigmetrics,probingScheme} for detailed studies of such effects.

Finally, in addition to the key advantage of providing homogeneous tree ego-centered views of the topology, the \tracetree\ tool also is much more efficient than \traceroute\ in terms of the number of packets sent, thus making it possible to repeatedly run it in 
radar measurements
with a reasonable network cost.


\section{Towards event detection.}
\label{sec_event}

One key interest of our measurements is that they make it possible to
observe the dynamics of the \ip{} internet topology from an ego-centered
perspective, at a time scale of a few minutes only.
In
particular, detecting {\em events} in this dynamics, \ie\ major
changes in the topology, is very appealing from a security and
modeling point of view.

A most natural direction to try and detect events is to observe the
number of distinct \ip\ addresses seen at {\em each} round, as plotted
in Figure~\ref{fig_events}. Clear events indeed appear in such plots,
under the form of downward peaks.
However, this provides little
information, if any: these peaks may be caused by temporary partial or total 
connectivity losses at the monitor (or close to it), not by important
events at the internet level. On the other hand, one may notice that
no significant upward peak appears in this plot. Notice that this is a
non-trivial fact: from a topological point of view, such peaks would
be possible; the fact that they do not occur reflects non-trivial
properties of the topology and its dynamics, which we leave for
further study.

Interestingly, the plot of the number of distinct \ip\ addresses seen
during ten consecutive rounds, Figure~\ref{fig_events}, has very
different characteristics. It exhibits
upward peaks (the distribution of observed values,  presented
in Figure~\ref{fig_cc}, left, confirms that these peaks are statistically significant
outliers). These peaks reveal important changes in the \ip\ addresses
observed in consecutive rounds, and thus important routing changes:
though the number of observed \ip{} addresses is roughly the same before
and after these events,
the ego-centered views have changed.

\begin{figure*}[!ht]
\begin{center}
\includegraphics[scale=0.5]{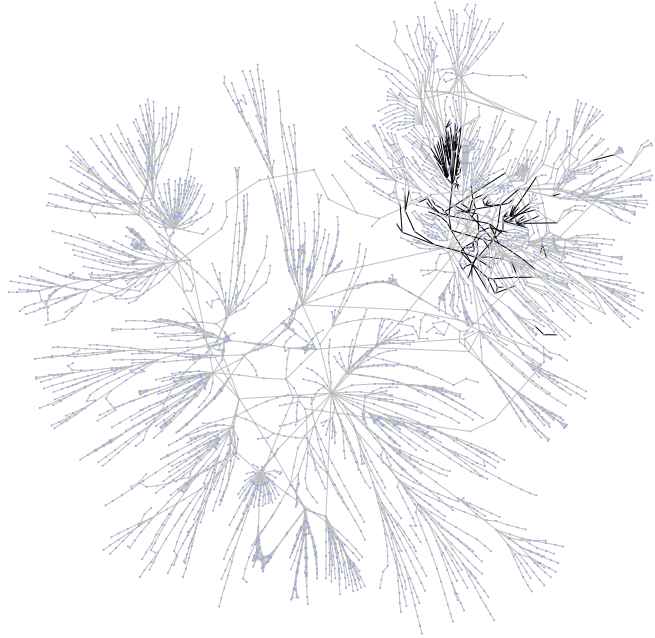}
\end{center}
\caption{Representation of the event at round 106231 in Figure~\ref{fig_events}:
the graph is obtained by merging 100 rounds before the event together with 
a single round after the event.
Edges in bold black are edges that were seen in the round after the event but not in the 
100 rounds before.
}
\label{fig_graph}
\end{figure*}

To illustrate this,
we present in Figure~\ref{fig_graph} a graph obtained by merging ego-centered
views measured before and after such an upward peak.
We can clearly see that this peak corresponds to a large number of new edges 
appearing in a specific part of the network,
confirming the occurrence of a significant event.

\medskip
Another approach consists in detecting
events occurring during a measurement from round $i$ to round $j$ by
comparing it to the measurement from round $i-k$ to round $i$, which
serves as a reference: we consider the \ip\ addresses seen during the
period of interest which were not observed in the reference period. We
call these \ip\ addresses the {\em new} addresses. 
Our observations show that it is natural to observe such new addresses
during any measurement.
However, one may expect
that events of interest will lead to the appearance of connected
groups of such addresses; we therefore propose to compute the
connected components composed of new addresses\,\footnote{\ie\ maximal
  sets of new addresses such that there exists a path between any two
  of them composed only of new addresses.} as a way to observe these
events.

We display such components in Figure~\ref{fig_cc} (center), together
with their neighborhood. This figure shows clearly that, in some
cases, the observed components are non-trivial islands of newly
observed nodes, revealing local events in the network. 
Figure~\ref{fig_cc} (right) however shows that such non-trivial islands are quite rare: most connected components of new nodes are very small, often reduced to a single node ($949$ over a total of $1\,457$ components, in our example). Despite this, some large components appear (the largest one in our example has size $17$, and $15$ components have size at least $10$), thus revealing underlying events of interest.

Another important characteristic of connected components of new addresses is the number of rounds needed to discover all their nodes, defined as the round number at which their last node was discovered minus the round number at which their first node was, plus one. Indeed, short discovery times indicate that all the new nodes under concern probably appeared because of a same event. Large times, instead, show that several events (located close to each other in the network) occurred. The examples in Figure~\ref{fig_cc} (center) show that both cases occur.

The distribution of the number of rounds needed to discover each component of new nodes (not represented here) is very heterogeneous, with many components discovered very rapidly and others much more slowly. This gives little information, however, as the discovery time may depend strongly on the component size. Studying the correlations between the two (not represented here) confirms this, but it also shows that some large connected components are discovered very rapidly.

\medskip

The two approaches we described point out specific moments at which events occurred;
one may then observe the data more
closely, in order to investigate the nature of these events. We leave
this for further research.

\section{Conclusion and perspectives.}
\label{sec_conclusion}

In this paper, we propose, implement, and illustrate a new measurement
approach which makes it possible to study the dynamics of \ip -level
internet topology at a time scale of a few minutes. We provide a rich
dataset consisting in radar measurements from more than one
hundred monitors towards thousands of destinations, conducted for
several weeks in continuous.

The most important direction for further research is of course the
analysis of collected data.
A particularly appealing goal is the detection of events
in the dynamics of the observed topology;
this raises difficult fundamental questions,
such as the characterization of {\em normal} dynamics,
or the identification of relevant time scales for the observation.

Other promising directions include visualizing the observed
dynamics,
and conducting more radar measurements to gain a deeper insight
(for instance, one could conduct simultaneous
measurements from several monitors to observe the dynamics from
different viewpoints).

\bigskip
\noindent
{\bf Acknowledgments.}
We warmly thank the PhD students and other colleagues of the LIP6, in particular Guillaume Valadon, Renata Teixeira,
and Brice Augustin who provided great insight during this work.
Likewise, we thank Beno\^\i t Donnet, who helped much with the references and also provided useful comments. Many interesting
discussions within the METROSEC project \cite{metrosecurl} also played a key role in our work.

We also thank all the people who provided monitors to us, in particular the PlanetLab staff \cite{planetLab}, Fr\'ed\'eric Aidouni,
Julien Aussibal, Prof. Hiroshi Esaki (WIDE), Jean-Charles de Longueville (Hellea) and S\'ebastien Wacquiez (Enix);
no such work would be possible without their help.

This work was funded in part by the METROSEC and AGRI projects.

\bibliographystyle{latex8}
\bibliography{anomalies,biblio,donnet}

\end{document}